\documentclass[reqno]{amsart}
\usepackage{amsmath,amsthm,amssymb}
\usepackage{latexsym}
\usepackage{eucal}

\newtheorem{theorem}{Theorem}[section]
\newtheorem{lemma}{Lemma}[section]

\newtheorem{Open question}{Open question}

\theoremstyle{definition}

\let\Im=\undefined
\DeclareMathOperator{\Im}{Im}

\begin{document}
\title[Wave propagation through sparse \ldots
\ldots]{Wave propagation through sparse potential barriers}
\author{Sergey A. Denisov}
\address{University of Wisconsin-Madison,
Mathematics Department, 480 Lincoln Dr., Madison, WI, 53706-1388,
USA,  {\rm e-mail: denissov@math.wisc.edu}} \maketitle

\begin{abstract}
We prove that 3-dimensional Schr\"odinger operator with slowly
decaying sparse potential has an a.c. spectrum that fills
$\mathbb{R^+}$. A new kind of WKB asymptotics for Green's function
is obtained. The absence of positive eigenvalues is established as
well.
\end{abstract} \vspace{1cm}

 Consider the Schr\"odinger operator
\begin{equation}
H=-\Delta+V, x\in \mathbb{R}^d \label{hamiltonian}
\end{equation}
We are interested in studying the scattering properties of $H$ for
the slowly decaying potential $V$. The following conjecture is due
to Barry Simon \cite{simon}

{\bf Conjecture.} If $V(x)$ is such that
\begin{equation}
\int\limits_{\mathbb{R}^d} \frac{V^2(x)}{1+|x|^{d-1}}\, dx<\infty
\label{barry}
\end{equation}
then $\sigma_{ac}(-\Delta+V)=\mathbb{R}^+$.

Some progress was recently made for slowly decaying oscillating
potentials \cite{lns, lns1, oleg, denisov} and potentials
asymptotically close to the spherically symmetric  \cite{perelman}.
But even for $V$ satisfying the bound $|V(x)|<C<x>^{-\gamma},$
$(\gamma<1, <x>=(1+|x|^2)^{0.5})$, there are no results. We also
want to mention that for the problem on the Bethe lattice (Caley
tree), we have relatively good understanding \cite{d1,d2}: consider
a rooted Caley tree  and denote the root by $O$. Assume that each
point has three neighbors and $O$ has only two. Consider the
discrete Schr\"odinger operator with potential $V$ and denote by
$d\sigma_O$ the spectral measure corresponding to the discrete delta
function at $O$. Let $w(\lambda)=(4\pi)^{-1} (8-\lambda^2)^{1/2}$ on
$[-2\sqrt 2, 2\sqrt 2]$, and
$\rho_O=\sigma'_O(\lambda)[w(\lambda)]^{-1}$, a relative density of
the spectral measure at the point $O$. Consider all paths that go
from $O$ to infinity without self-intersections and define the
probability space on the set of these paths by assigning to each of
them the same weight (i.e. as we go from $O$ to infinity, we toss
the coin at any vertex and move to one of the neighbors further from
$O$ depending on the result). In \cite{d2}, we proved the following
\begin{theorem} For any bounded $V$, define
\[
s_O=\mathop\int\limits_{-2\sqrt 2}^{2\sqrt 2}  \ln
 \rho_O(\lambda) w(\lambda) d\lambda
\]
Then the following inequality is true
\begin{equation}
\exp s_O\geq  \mathbb{E} \left\{ \exp \left[-\frac 14
\sum\limits_{n=1}^\infty V^2(x_n)\right]\right\}\label{pearson}
\end{equation}
where the expectation is taken with respect to all paths $\{x_n\}$
going from $O$ to infinity without self-intersections. In
particular, if the r.h.s. of (\ref{pearson}) is positive, then
$[-2\sqrt 2,2\sqrt 2]\subseteq \sigma_{ac}(H)$. \label{improve}
\end{theorem}
Notice that for the r.h.s. to be positive we just need to make sure
that there are ``enough" paths over which the potential is square
summable. This is much weaker than (\ref{barry}).

In the current paper, we consider the following model. Let $R_n$ be
a sparse sequence of real numbers, i.e. $R_n\to+\infty,
R_{n+1}/R_n\to\infty$ as $n\to\infty$.  Consider concentric
spherical layers $\Sigma_n=\{x: R_n<|x|<R_n+1\}$ and assume that the
measurable functions $v_n(x)=0$ outside these layers and
$|v_n(x)|<v_n, x\in \Sigma_n$, $v_n\to 0$ as $n\to\infty$. Take
\[
V_n(x)=\sum\limits_{j=1}^n v_j(x), V(x)=\sum\limits_{j=1}^\infty
v_j(x)
\]
We will study the scattering properties of the corresponding $H$. In
particular, we want to study the a.c. spectrum and the spatial
asymptotics of the Green function. The sparse potentials of general
 form were studied earlier (see, e.g. \cite{molchanov, mv, krishna} and
references there).  It is also necessary to mention the
one-dimensional result first (see  \cite{pearson,kls}).
\begin{theorem}(see \cite{kls})
Assume that $d=1$, $R_n/R_{n+1}\to 0$ and $v_n\to 0$. Then for
\[
V(x)=\sum_{n=1}^\infty v_n \phi(x-R_n)
\]
($\phi(x)$--a nonzero, nonnegative bump function), we have the
following: if $v_n\in \ell^p(\mathbb{Z}^+), p\leq 2$, the spectrum
of $H$ is purely a.c. on $\mathbb{R}^+$. For $p>2$, it is singular
continuous.
\end{theorem}

We are interested in studying the same phenomena in the
multidimensional case. Clearly, by taking the potential spherically
symmetric, one can show that the spectrum can be singular continuous
for $v_n\in \ell^p, p>2$ even in multidimensional case. We will
address the following question: assume that the sequence $\{R_n\}$
is as sparse as we like and $v_n$ is an arbitrary sequence from \
$\ell^2(\mathbb{Z}^+)$. Is it true that the a.c. spectrum exists?

  From Weyl's theorem, we know that $\sigma_{ess}(H)=[0,\infty)$.
We need to introduce certain quantities. Let $z=k^2$ and $\Pi=\{k\in
\mathbb{C}^+, 0<\Im k<1\}$. If $f(x)$ is nonzero $L^2$ function with
compact support, say, within the unit ball, then
$u_n(x,k)=(-\Delta+V_n-k^2)^{-1}f$ has the following asymptotics at
infinity

\begin{equation}
\begin{array}{cc}
\displaystyle u_n(x,k)=&\displaystyle \frac{\exp (ikr)}{r}\left(
A_n(f,k,\theta)+\bar{o}(1)\right), \\
\displaystyle \frac{\partial u_n(x,k)}{\partial r}= &ik
\displaystyle \frac{\exp
(ikr)}{r}\left( A_n(f,k,\theta)+\bar{o}(1)\right),\\
\displaystyle & r=|x|, \theta=\displaystyle \frac{x}{|x|},
|x|\to\infty
 \end{array}
 \quad {\rm (Sommerfeld's\; radiation\; conditions)}
\end{equation}
as long as $k\in \mathbb{C}^+$ and $k^2\notin \sigma(H_n)$. Since
$V_n$ has compact support, one can show that $A_n$ is continuous in
$k$ up to the real line. The physical meaning of $A_n(f,k,\theta)$
is the amplitude of the outgoing spherical wave in the direction
$\theta$ after propagation through $n$ concentric  barriers. For the
free case (i.e. $V=0$) the amplitude of function $f$ is given by the
formula
\[
A^0(f,k,\theta)=(4\pi)^{-1}\int\limits_{\mathbb{R}^3}
\exp(-ik<\theta,x>) f(x)dx
\]
Notice that for the fixed $\theta$ this function is entire in $k$
(nonzero $f$ has compact support) and so has finite number of zeroes
inside any compact.

Instead of proving the asymptotics of the Green function of $H$, we
will study an asymptotical behavior of the sequence
$A_n(f,k,\theta)$, as $k\in \Pi$, and $n\to\infty$. That will
contain all relevant information on the scattering mechanism. The
following is the main result of the paper

\begin{theorem}
If the sequence $R_n$ is sparse enough{\footnote{For example, an
estimate $R_{n+1}>e^{\alpha R_n}, \alpha>1, n\in \mathbb{Z}^+$ will
be sufficient}}, then for any $v_n\in \ell^2(\mathbb{Z}^+)$, we have

\begin{equation}
A_n(f,k,\theta)=WKB_n(k,\theta) \tilde{A}_n(f,k,\theta)
\label{asymp1}
\end{equation}
where \begin{equation} WKB_n(k,\theta)=
\exp\left[-(4\pi)^{-1}\int\limits_{\mathbb{R}^3}
\frac{\exp(ik(|t|-<\theta, t>))}{|t|} V_n(t)dt\right],
|\tilde{A}_n(f,k,\theta)|<C(k) \label{asymp2}
\end{equation} uniformly in $n$ for
 $k: k=\tau+i\epsilon, \tau>0, 0<\epsilon<1$. If $k$ is fixed, then
\begin{equation}
\tilde{A}_n(f,k,\theta)=A^0(f,k,\theta)+\delta(\|v\|_2,k)
\label{asymp3}
\end{equation}
where $\delta(\|v\|_2,k)\to 0$ as $\|v\|_2\to 0$ uniformly in $n$.
Moreover, $\sigma_{ac}(H)=\mathbb{R}^+$.\label{theorem1}

\end{theorem}

\begin{proof} For simplicity, assume $\|v\|_2<1$.
From now on, we will reserve the symbol $C$ for the constant whose
value can change from one formula to another. We need two lemmas
first.

\begin{lemma}
Consider potential $V(x): V(x)=0$ for $|x|>R$ and $|V(x)|<1$. Denote
the Green function by $G(x,y,k)$. Then, for $k: k=\tau+i\epsilon;
\tau, \epsilon>0$, we have

\begin{equation}
G(x,y,k)=\frac{e^{ik|x-y|}}{4\pi|x-y|}+\delta(x,y,k)
\end{equation}
where
\begin{equation}
|\delta(x,y,k)|\leq C\frac{R^3 e^{2\epsilon R-\epsilon |x|-\epsilon
|y|}}{\tau \epsilon (|x|-R)(|y|-R)}; |x|,|y|>>R \label{est1}
\end{equation}
Moreover, if
\[
B(\hat{x}, y, k)=\lim\limits_{|x|\to\infty} \left(|x|e^{-ik|x|}
G(x,y,k)\right)
\]
then
\begin{equation}
|\partial B(\hat{x}, y,k)|<(4\pi)^{-1}|k||y|\exp(\epsilon <\hat{x},
y>) +CR|k|\left(\frac{R^3 e^{2\epsilon R-\epsilon |y|)}}{\tau
\epsilon (|y|-R)}\right), \hat{x}=x/|x| \label{est2}
\end{equation}
and the derivative is taken with respect to $\hat{x}\in \Sigma$ (a
unit sphere).
 \label{lemma1}
\end{lemma}

\begin{proof}
By the second resolvent identity,
\[
G(x,y,k)=(4\pi)^{-1}\frac{e^{ik|x-y|}}{|x-y|}-(4\pi)^{-2}\int\limits
\frac{e^{ik|x-u|}}{|x-u|}V(u)\frac{e^{ik|u-y|}}{|u-y|}du\]
\[
+(4\pi)^{-2}\int\limits \frac{e^{ik|x-u|}}{|x-u|} V(u)\int\limits
G(u,t,k) V(t) \frac{e^{ik|t-y|}}{|t-y|} dtdu
\]
For the second term,
\[
\left|\int\limits
\frac{e^{ik|x-u|}}{|x-u|}V(u)\frac{e^{ik|u-y|}}{|u-y|}du\right|<
C\frac{R^3 e^{2\epsilon R-\epsilon |x|-\epsilon |y|}}{
(|x|-R)(|y|-R)}
\]
To estimate the third term, we notice that $\Im k^2=2\tau\epsilon$
and therefore
\[
\left|\int\limits \frac{e^{ik|x-u|}}{|x-u|} V(u)\int\limits G(u,t,k)
V(t) \frac{e^{ik|t-y|}}{|t-y|} dtdu\right|\] \[<
(2\tau\epsilon)^{-1} \left( \,\int\limits_{|u|<R}
\frac{e^{-2\epsilon |x-u|}}{|x-u|^2} du \right)^{1/2} \left(\,
\int\limits_{|t|<R} \frac{e^{-2\epsilon |t-y|}}{|t-y|^2} dt
\right)^{1/2}
\]
Now (\ref{est1}) is straightforward.

To obtain (\ref{est2}), we write the following bounds for any unit
vector $\nu\in\mathbb{T}_{\hat{x}}$,
\[
|\partial_\nu B(\hat{x}, y,k)|< (4\pi)^{-1}|k<\nu, y>
e^{-ik<\hat{x}, y>}| +(4\pi)^{-2}\left| \int\limits k <\nu, u>
e^{-ik<\hat{x}, u>}V(u)\frac{e^{ik|u-y|}}{|u-y|}du\right|\]
\[
+(4\pi)^{-2}\left|\int\limits k<\nu, u> e^{-ik<\hat{x}, u>}
V(u)\int\limits G(u,t,k) V(t) \frac{e^{ik|t-y|}}{|t-y|} dtdu\right|
\]
Then, we estimate the second and the third terms as before.
\end{proof}

\begin{lemma}
Under the conditions of lemma \ref{lemma1}, consider nonzero
$f(x)\in L^2$ with compact support (say, inside the unit ball).
Define the corresponding amplitude $A(f,k,\hat{x})$. Then,
\begin{equation}
u(x,k)=(H-k^2)^{-1}f=\frac{\exp(ik|x|)}{|x|}\left(
A(f,k,\hat{x})+\rho(x,k)\right)
\end{equation}
where
\begin{equation}
|\rho(x,k)|<C(\tau\epsilon)^{-1} \frac{R^{3.5}e^{\epsilon
R}}{|x|-R}; |x|>>R \label{est3}
\end{equation} \label{lemma2}
\end{lemma}
\begin{proof}
We have $-\Delta u+Vu=f+k^2u$ which can be rewritten
$u=(-\Delta-k^2)^{-1}\mu$, where $\mu=f-Vu$ has compact support. So,
\[
u(x,k)=(4\pi)^{-1}\int\limits\frac{e^{ik|x-t|}}{|x-t|} \mu(t)dt
\]
Clearly, \[ A(f,k,\hat{x})=(4\pi)^{-1}\int e^{-ik<\hat{x},
t>}\mu(t)dt\] and
\[
\int\left|\left(\frac{|x|}{|x-t|} e^{ik(|x-t|-|x|)}-e^{-ik<\hat{x},
t>}\right)\mu(t)\right|dt
\]
\[
< \|\mu\|\left[ \int\limits_{|t|<R}  \left|\frac{|x|}{|x-t|}
e^{ik(|x-t|-|x|)}-e^{-ik<\hat{x}, t>}\right|^2dt \right]^{1/2}
\]
\[
<C(\tau \epsilon)^{-1} R^{1.5} \sup_{|t|<R}
\left|\left(\frac{|x|}{|x-t|}-1\right)
e^{ik(|x-t|-|x|)}+(e^{ik|x-t|-ik|x|}-e^{-ik<\hat{x}, t>})\right|
\]
Then, the obvious estimates lead to (\ref{est3}).

\end{proof}

Now, let us proceed to the proof of the theorem. The strategy is
rather simple. We want to obtain recursion for $A_{n}$. To do that,
we will consecutively perturb $H_{n}$ by $v_{n+1}(x)$, $H_{n+1}$ by
$v_{n+2}(x)$, etc. That will allow us to obtain almost
multiplicative representation for the amplitude, so well-known in
the one-dimensional case. On the level of physical intuition, by
requiring the sparseness of barriers, we make sure that the wave,
propagated through $n$ barriers, hits the $n+1$ barrier almost like
an outgoing spherical wave. That makes an analysis doable.

Let us write the second resolvent identity for $H_{n+1}=-\Delta
+V_n(x)+v_{n+1}(x)$:

\begin{equation}
G_{n+1}(x,y,k)=G_n(x,y,k)-\int G_n(x,u,k)v_{n+1}(u)G_n(u,y,k)du
\end{equation}
\[
+\int G_n(x,u,k)v_{n+1}(u)\int G_{n+1}(u,s,k) v_{n+1}(s)
G_n(s,y,k)ds
\]
Therefore,
\[
u_{n+1}(x)=\int G_{n+1}(x,y,k) f(y)dy=u_{n}(x)-\int
G_n(x,y,k)v_{n+1}(y)u_n(y)dy
\]
\[
+\int G_n(x,y,k)v_{n+1}(y)\int G_{n+1}(y,s,k) v_{n+1}(s) u_n(s)dsdy
\]
Taking $|x|$ to infinity, we get

\begin{equation}
A_{n+1}(k,\hat{x})=A_{n}(k,\hat{x})-(4\pi)^{-1}\int
e^{-ik<\hat{x},u>} v_{n+1}(u)
\frac{e^{ik|u|}}{|u|}A_n(k,\hat{u})du+r_n(k,\hat{x}) \label{recur1}
\end{equation}
Introducing the spherical variables, we see that the second term is
not greater than
\begin{equation}C\|A_n\|_\infty
v_{n+1}\epsilon^{-1}\label{second}\end{equation}
 We will see that $r_n$ can be regarded as a small correction
to the recurrence relation which basically looks like this:
\begin{equation}
l_{n+1}=\left[I-\int\limits_{R_{n+1}}^{R_{n+1}+1}O_{t}q_{t}dt\right]l_{n}
\label{iter}
\end{equation}
where $l_n(\theta)$ are the functions on $\Sigma$,
\begin{equation}
O_t f(\theta)=(4\pi)^{-1}t\int_{\Sigma} e^{ikt(1-<\theta,s>)} f(s)ds
\label{ooo}
\end{equation}
and $q_t$ is just an operator of multiplication by the function
$q_t(\theta)$ given on the unit sphere. The difficulty of the
problem comes from noncommutativity of $O_n$ and $q_n$. If not the
correction $r_n$ we would just have the product of operators. But to
get the needed asymptotics for this product, we will have to
essentially use the sparseness condition again.

The rest of the proof goes as follows: we first obtain rough apriori
estimates on $A_n(k,\theta)$ and $\partial A_n(k,\theta)$. Then, we
will use them to obtain an accurate asymptotics for $A_n(k,\theta)$.
In the last part, this asymptotics will be used to show the presence
of a.c. component of the spectral measure.

For $r_n$, we have $r_n=I_1+\ldots+I_7$. Applying lemma \ref{lemma1}
and lemma \ref{lemma2}, we get the following estimates for $I_j$:

\[
I_1= -\lim_{|x|\to\infty} |x|e^{-ik|x|} \int
\frac{e^{ik|x-u|}}{4\pi|x-u|}v_{n+1}(u)\frac{e^{ik|u|}}{|u|}\rho_n(u,k)du
\]

\[
|I_1|<C \left|\int e^{-ik<\hat{x},u>} v_{n+1}(u)
\frac{e^{ik|u|}}{|u|}\rho_n(u,k)du\right|
\]
\[
< \sigma_n (\tau\epsilon)^{-1} v_{n+1}
\int\limits_{R_{n+1}<|u|<R_{n+1}+1} \frac{e^{\epsilon(<\hat{x},
u>-|u|)}}{|u|}du
\]
where
\begin{equation}\sigma_n=R_n^{3.5}e^{R_n}(R_{n+1}-R_n)^{-1}\end{equation} By
introducing the spherical coordinates, we estimate the last integral
\[
\int\limits_{R_{n+1}<|u|<R_{n+1}+1} \frac{e^{\epsilon(<\hat{x},
u>-|u|)}}{|u|}du=C\int\limits_{R_{n+1}}^{R_{n+1}+1} \rho
\frac{1-e^{-2\epsilon\rho}}{\epsilon \rho} d \rho<C\epsilon^{-1}
\]
Thus, \begin{equation}|I_1|<C(\tau\epsilon^2)^{-1}\sigma_n
v_{n+1}\label{I1}\end{equation}

For $I_2$:

\[
I_2=-\lim_{|x|\to\infty} |x|e^{-ik|x|} \int \delta_n(x,t,k)
v_{n+1}(t)\frac{e^{ik|t|}}{|t|}A_n(\hat{t},k)dt
\]

\[
|I_2|<C \int \frac{R_n^3 e^{2\epsilon R_n-\epsilon
|t|}}{\tau\epsilon (t-R_n)} v_{n+1}(t) \frac{e^{-\epsilon |t|}}{|t|}
|A_n(\hat{t},k)|dt
\]
\begin{equation}
<C(\tau\epsilon)^{-1}R_n^3 v_{n+1} e^{2\epsilon(R_n-R_{n+1})}
\|A_n(\hat{t}, k)\|_{L^\infty(\Sigma)}\label{I2}
\end{equation}
as long as
\begin{equation}
R_{n+1}>2R_n \label{sparse2}
\end{equation}

Define $I_3$ as
\[
I_3=-\lim_{|x|\to\infty} |x|e^{-ik|x|} \int \delta_n(x,t,k)
v_{n+1}(t)\frac{e^{ik|t|}}{|t|}\rho_n(t,k)dt
\]

\begin{equation}
|I_3|< C(\tau\epsilon)^{-1} R_n^3 \sigma_n \int \frac{e^{2\epsilon
R_n-2\epsilon |t|}}{\tau \epsilon (|t|-R_n)
|t|}|v_{n+1}(t)|dt<C\tau^{-2}\epsilon^{-2} v_{n+1} \sigma_n R_{n}^3
e^{2\epsilon (R_n-R_{n+1})}\label{I3}
\end{equation}

For the other terms, we will be using the so-called Combes-Thomas
inequality \cite{klein}, which says the following. Assume that
potential $Q$ is bounded. Then, for the {\it operator kernel}, we
have \footnote{The actual estimate obtained in \cite{klein} is
stronger}
\[\|\chi(x) (-\Delta+Q-z)^{-1} \chi(x)\|<C(\Im z)^{-1}
e^{-\gamma \Im z |x-y|}
\]
where $\chi(x)$ is characteristic function of the unit cube centered
at $x$, $\gamma$ is fixed positive parameter and the norm is
understood as the norm of operator acting in $L^2(\mathbb{R}^3)$.
Although an estimate on the operator kernel is not the same as
pointwise estimate on the Green function, it is almost the same in
our case. Let us accurately show that for $I_4$, for the other
terms, we will be skipping details. We have

\[
I_4=\lim_{|x|\to\infty} |x|e^{-ik|x|} \int
\frac{e^{ik|x-u|}}{4\pi|x-u|}v_{n+1}(u) \int
G_{n+1}(u,s,k)v_{n+1}(s) \frac{e^{ik|s|}}{|s|}A_n(\hat{s},k)dy
\]

\[
|I_4|< C\int e^{\epsilon <\hat{x},u>} |v_{n+1}(u)|\left|\int
G_{n+1}(u,s,k)v_{n+1}(s)
\frac{e^{ik|s|}}{|s|}A_n(\hat{s},k)ds\right|du
\]
\begin{equation}
<\sum\limits_{u_i, s_j} \int\limits_{C(u_i)} e^{\epsilon
<\hat{x},u>} |v_{n+1}(u)|\left|\,\int\limits_{C(s_j)}
G_{n+1}(u,s,k)v_{n+1}(s)
\frac{e^{ik|s|}}{|s|}A_n(\hat{s},k)ds\right|du \label{I4}
\end{equation}
where $C(u_i)$ are all unit cubes from the $\mathbb{Z}^3$ partition
of $\mathbb{R}^3$ that intersect $\Sigma_{n+1}$. Points $u_i$ (same
as $s_j$) are the centers of these cubes. Therefore, we have

\[
|I_4|< Cv_{n+1}^2 \|A_n\|_\infty \sum\limits_{u_i,s_j} e^{\epsilon
<\hat{x},u_i>} \frac{e^{-\gamma \epsilon \tau |u_i-s_j|}}{\tau
\epsilon} \frac{e^{-\epsilon |s_j|}}{|s_j|}
\]
\[
<Cv_{n+1}^2 \|A_n\|_\infty \int\limits_{R_{n+1}-2<|u|<R_{n+1}+3}
e^{\epsilon <\hat{x},u>} \int\limits_{R_{n+1}-2<|s|<R_{n+1}+3}
\frac{e^{-\gamma \tau \epsilon |u-s|}}{\tau \epsilon}
\frac{e^{-\epsilon |s|}}{|s|}dsdu
\]
So, \begin{equation} |I_4|<Cv_{n+1}^2
\|A_n\|_\infty\tau^{-4}\epsilon^{-5} \label{I4}\end{equation} In the
same way, we have
\[
I_5=\lim_{|x|\to\infty} |x|e^{-ik|x|} \int \delta_n(x,y,k)
v_{n+1}(y) \int G_{n+1}(y,s,k)v_{n+1}(s)
\frac{e^{ik|s|}}{|s|}A_n(\hat{s},k)dy
\]

\[
|I_5|<C\int \frac{R_n^3 e^{2\epsilon R_n-\epsilon |y|}}{\tau\epsilon
(|y|-R_n)} |v_{n+1}(y)| \left| \int G_{n+1} (y,s,k) v_{n+1}(s)
\frac{e^{ik|s|}}{|s|}A_n(\hat{s},k)ds\right|dy
\]
\begin{equation}
<C(\tau\epsilon)^{-5} R_n^3 v_{n+1}^2 \|A_n\|_\infty e^{2\epsilon
(R_n-R_{n+1})} \label{I5}
\end{equation}

For $I_6$:

\[
I_6=\lim_{|x|\to\infty} |x|e^{-ik|x|} \int \delta_n(x,y,k)
v_{n+1}(y) \int G_{n+1}(y,s,k)v_{n+1}(s)
\frac{e^{ik|s|}}{|s|}\rho_n(s,k)dy
\]

\begin{equation} |I_6|<C(\tau\epsilon)^{-6} R_n^3 \sigma_n v_{n+1}^2
e^{2\epsilon (R_n-R_{n+1})} \label{I6}
\end{equation}

Define $I_7$:

\[
I_7= \lim_{|x|\to\infty} |x|e^{-ik|x|} \int
\frac{e^{ik|x-y|}}{4\pi|x-y|}v_{n+1}(y) \int
G_{n+1}(y,s,k)v_{n+1}(s) \frac{e^{ik|s|}}{|s|}\rho_n(s,k)dy
\]

\begin{equation}
|I_7|<C(\tau\epsilon)^{-6}\sigma_n v_{n+1}^2 \label{I7}
\end{equation}
From now on we assume that $\tau$ changes within the interval
$I=[a,b], a>0$. Then, we can disregard dependence on $\tau $ and
will keep track on $\epsilon$ only. The estimates on $I_j$ and
(\ref{second}) amount to
\[
\|A_{n+1}\|_\infty<
\|A_n\|\left(1+Cv_{n+1}\epsilon^{-5}+Cv_{n+1}\epsilon^{-5} R_n^3
e^{2\epsilon(R_n-R_{n+1})}\right)
\]

\begin{equation}
+C\epsilon^{-6} \sigma_n v_{n+1} +C\epsilon^{-6} \sigma_n R_n^3
v_{n+1} e^{2\epsilon (R_n-R_{n+1})}\label{recur2}
\end{equation}

The following lemma is trivial

\begin{lemma}
If $x_{n}, a_n, b_n\geq 0$ and $x_{n+1}\leq a_nx_n+b_n$, then
\[
x_{n+1}\leq (x_0+\sum\limits_{j=0}^n b_j)\max\limits_{j=0,1,\ldots,
n}\{1, a_j\cdot a_{j+1}\cdot\ldots \cdot a_{n-1} \cdot a_n\}
\]\label{lemma3}
\end{lemma}
\begin{proof}
The proof follows from the iteration of the given inequality.
\end{proof}

\begin{lemma}
The following estimates hold
\[
x^je^{-\epsilon x}\leq (j/e)^j\epsilon^{-j}
\]
for any $x>0, j>0, \epsilon>0$.\label{lemma4}
\end{lemma}
\begin{proof}
The function $f(x)=x^je^{-\epsilon x}$ has maximum at the point
$x^*=j\epsilon^{-1}$.
\end{proof}

By using lemma \ref{lemma4} and estimate (\ref{sparse2}), we get
\begin{equation}
\|A_{n+1}\|_\infty< \|A_n\|\left(1+Cv_{n+1}\epsilon^{-8}\right)
+C\epsilon^{-9} \sigma_n v_{n+1} \label{recur3}
\end{equation}

From lemma \ref{lemma3}, it follows
\begin{equation}
\|A_n\|_\infty<(\|A_0\|_\infty+\epsilon^{-9}\sum\limits_{j=0}^{n-1}
v_{j+1}\sigma_j )\exp(C\epsilon^{-8}\|v\|_2 n^{0.5})
\end{equation}
Let us make another, rather strong,  assumption on sparseness of
$R_n$:

\begin{equation}
\sigma_n<e^{-n}\label{sparse3}
\end{equation}
Then we have the following apriori bound on $\|A_n\|_\infty$:

\begin{equation}
\|A_n\|_\infty<g_n=(\|A_0\|_\infty+C\|v\|_2\epsilon^{-9})\exp(C\epsilon^{-8}\|v\|_2
n^{0.5}) \label{apriori1}
\end{equation}
We will need analogous estimate for the derivatives of $A_n$. Let us
use formula (\ref{recur1}) and the same bounds as before together
with estimate (\ref{est2}). From (\ref{est2}), we see that we only
pick up extra $R_{n+1}$ in the inequalities. So,

\[
|\partial A_{n+1}|< |\partial A_n|+
C\|A_n\|R_{n+1}v_{n+1}\epsilon^{-8} +C\epsilon^{-9} R_{n+1}\sigma_n
v_{n+1}
\]
and from (\ref{apriori1}) we get
\[
\|\partial A_{n+1}\|_\infty<g'_{n+1}=\|\partial A_0\|+R_{n+1}
\epsilon^{-9}\|v\|_2+ \]
\begin{equation}
+nR_{n+1}\|v\|_2\epsilon^{-8} \exp(C\epsilon^{-8}\|v\|_2 n^{0.5})
\left(\|A_0\|_\infty+ C\|v\|_2 \epsilon^{-9}\right)\label{apriori2}
\end{equation}
Now we are going to use these apriori estimates to obtain
asymptotics of the sequence $A_n$. By considering $R_1$ big enough,
we may assume that $A_0$ is the amplitude for the unperturbed
operator. Consider the second term in (\ref{recur1}). It can be
written as
\[
(4\pi)^{-1}A_n(k,\hat{x}) \int
\frac{v_{n+1}(u)}{|u|}e^{-ik<\hat{x},u>+ik|u|}du
\]
\[
 +(4\pi)^{-1}\int \frac{v_{n+1}(u)}{|u|}(A_n(k,\hat{u})-A_n(k,\hat{x}))e^{-ik<\hat{x},u>+ik|u|}du
 \]
Denote \[\kappa_n=(4\pi)^{-1}\int
\frac{v_{n+1}(u)}{|u|}e^{-ik<\hat{x},u>+ik|u|}du\] Notice that
\begin{equation}
|\kappa_n|<C\epsilon^{-1}v_{n+1}\in \ell^2(\mathbb{Z}^+)
\label{kappan}
\end{equation}
The second term can be bounded using an apriori bound on the
derivative. It is not greater than
\[
C\|\partial A_n\|_\infty\int e^{-\epsilon
(|u|-<\hat{x},u>)}|v_{n+1}(u)| \frac{|\hat{u}-\hat{x}|}{|u|} du
\]
\[
<Cg'_nv_{n+1}R_{n+1}\left[\int\limits_{\pi/2-\delta}^{\pi/2}
e^{-\epsilon R_{n+1}(1-\sin\theta)}\cos^2\theta
d\theta+\int\limits_{-\pi/2}^{\pi/2-\delta} e^{-\epsilon
R_{n+1}(1-\sin\theta)}\cos\theta d\theta\right]
\]
\[
<Cg'_nv_{n+1}R_{n+1}\left[ \epsilon^{-1.5}R_{n+1}^{-1.5}
+\epsilon^{-1}R_{n+1}^{-1}e^{-\delta_1\epsilon R_{n+1}} \right]
\]
\begin{equation}
<Cg'_nv_{n+1}R_{n+1}\left[ \epsilon^{-1.5}R_{n+1}^{-1.5}
+\epsilon^{-2}R_{n+1}^{-2} \right]\label{est-ax}\end{equation}  and
used lemma \ref{lemma4} in the last inequality. Now, we will pay
special attention to $I_4$. The other terms will be of little
importance. We have

\[
I_4=\beta_n A_n(k,\hat{x})
\]
\begin{equation}
+ (4\pi)^{-1}\int e^{-ik<\hat{x},u>} v_{n+1}(u) \int G_{n+1}(u,s,k)
v_{n+1}(s) \frac{e^{ik|s|}}{|s|} (A_n(k,\hat{s})-A_n(k,\hat{x}))dsdu
\end{equation}
where
\[
\beta_n=(4\pi)^{-1}\int e^{-ik<\hat{x},u>} v_{n+1}(u) \int
G_{n+1}(u,s,k) v_{n+1}(s) \frac{e^{ik|s|}}{|s|}dsdu
\]
and upon using Combes-Thomas estimate, we have
\begin{equation}
|\beta_{n}|< C\epsilon^{-5}v_{n+1}^2 \label{betan}
\end{equation}
and it is very important that $\beta_n\in \ell^1({\mathbb{Z}^+})$.
That is the only place where we essentially use $\ell^2$ condition
on $v$. The other term can be estimated by

\[
C \epsilon^{-1} v_{n+1}^2 \|\partial A_n\|_\infty
\int\limits_{R_{n+1}-2<|u|<R_{n+1}+3} e^{\epsilon <\hat{x},u>}
\]
\begin{equation}
\times \int\limits_{R_{n+1}-2<|s|<R_{n+1}+3} e^{-\gamma \epsilon
|u-s|} \frac{e^{-\epsilon |s|}}{|s|}
(|\hat{s}-\hat{u}|+|\hat{u}-\hat{x}|)dsdu \label{terms}
\end{equation}
The first term in the last expression is bounded by
\[
Cv_{n+1}^2 \epsilon^{-2} g_n' R_{n+1}^2 \int_{\Sigma} e^{-\gamma
\epsilon R_{n+1}|\hat{u}-\hat{s}|} |\hat{u}-\hat{s}| ds
\]
\[
<Cv_{n+1}^2 \epsilon^{-2} g_n' R_{n+1}^2 \left[\,
\int\limits_{\pi/2-\delta}^{\pi/2} e^{-\gamma \epsilon R_{n+1} \cos
\theta }\cos^2 \theta d\theta+
e^{-\gamma\epsilon\delta_1R_{n+1}}\right]
\]
\begin{equation}
<Cv_{n+1}^2 \epsilon^{-5} g_n' R_{n+1}^{-1} \label{term1}
\end{equation}
The second term in (\ref{terms}) is estimated by
\[
C\epsilon^{-1}v_{n+1}^2 g_n' \int_{R_{n+1}-2<|u|<R_{n+1}+3}
e^{\epsilon <\hat{x},u>} |\hat{u}-\hat{x}|
\int_{R_{n+1}-2<|s|<R_{n+1}+3} e^{-\gamma\epsilon |u-s|}
\frac{e^{-\epsilon|s|}}{|s|} dsdu
\]
\[
< C\epsilon^{-4} v_{n+1}^2 g_n' R_{n+1}\left[\,
\int\limits_{\pi/2-\delta}^{\pi/2}  e^{-\epsilon R_{n+1}(1-\sin
\theta)} \cos^2\theta d\theta  + \int\limits_{-\pi/2}^{\pi/2-\delta}
e^{-\epsilon R_{n+1} (1-\sin\theta)} \cos\theta d\theta \right]
\]
\begin{equation}
<C\epsilon^{-6} v_{n+1}^2 g_n'  R_{n+1}^{-0.5} \label{term2}
\end{equation}
For the other $I_j$, we are using estimates obtained before and get
the following recursion

\begin{equation}
A_{n+1}(k,\theta)=A_n(k,\theta)(1-\kappa_n+\beta_n)+\eta_n
\label{recur4}
\end{equation}
 For $\eta_n$, we apply estimates
(\ref{I1}),(\ref{I2}),(\ref{I3}),(\ref{I5}),(\ref{I6}),(\ref{I7}),(\ref{est-ax}),(\ref{terms}),(\ref{term1}),(\ref{term2})
to get
\begin{equation}
|\eta_n|< C\epsilon^{-d} v_{n+1}\left\{
g_n'R_{n+1}^{-0.5}+\sigma_n+R_n^3g_ne^{2\epsilon(R_n-R_{n+1})}\right\}
\end{equation}
with some $d$ that will be reserved for the positive constant that
might change from formula to formula, we will not care for its
particular value. From the estimates on $g_n$ and $g'_{n}$, we get
\[
|\eta_n|<C\epsilon^{-d} v_{n+1}
\left[e^{C\epsilon^{-8}n^{0.5}}\left\{ nR_n
R_{n+1}^{-0.5}+R_n^3e^{2\epsilon(R_n-R_{n+1})}\right\}+\sigma_n\right]
\]
Assuming \begin{equation} nR_n R_{n+1}^{-0.5}<e^{-2n}, 2R_n<
R_{n+1}, \sigma_n<e^{-n}
\end{equation} we have
\begin{equation}
|\eta_n|<C\epsilon^{-d}v_{n+1}\exp(C\epsilon^{-8}n^{0.5}-1.5\epsilon
n)<\exp(C\epsilon^{-d})v_{n+1}\exp(- \epsilon n) \label{est-res}
\end{equation}

Now, we are going to use the following lemma
\begin{lemma}
Assume that $x_{n+1}=x_n(1+q_n)+d_n$ where $q_n\in
\ell^2(\mathbb{Z}^+)$, $|d_n|<\omega  e^{-\alpha n}, \alpha>0$, and
$n=0,1,\ldots$. Then, we have
\begin{equation}
x_n=\underline{O}\left[\exp(C\|q\|_2^2)\right] \exp\left[
\sum\limits_{j=0}^{n-1} q_j\right] \left( x_0 + \omega\sum_{j=1}^n
\exp(-\alpha j +\|q\|_2 j^{0.5})\right] \label{drob}
\end{equation} \label{lemma5}
\end{lemma}
\begin{proof}
Iterating, we get
\[
x_{n+1}=(1+q_n)\ldots(1+q_0)x_0+(1+q_n)\ldots(1+q_1)d_0+ \ldots +
(1+q_n)d_{n-1}+d_n
\]
Clearly,
\[
(1+q_n)\ldots(1+q_k)=\exp\left[ \sum\limits_{j=0}^n q_j\right]
r_{k,n}
\]
where
\[
r_{k,n}=\exp\left[- \sum\limits_{j=0}^{k-1}
q_j\right]\prod\limits_{j=k}^n l(q_j)
\]
and
\[
l(z)=(1+z)e^{-z}
\]
It is obvious that $|l(z)|<\exp\left[C|z|^2\right]$.  Then,
\[
\left|\exp\left[- \sum\limits_{j=0}^{k-1} q_j\right]\right|<
\exp\left[\sqrt k \|q\|_2\right]
\]
and
\[
\prod\limits_{j=k}^n |l(q_j)|<\exp\left[C\|q\|_2^2\right]
\]
Now, (\ref{drob}) easily follows.
\end{proof}

Let us now apply this lemma to (\ref{recur4}) bearing in mind
(\ref{est-res}), (\ref{kappan}), (\ref{betan}).  Then, for $R_0$
large enough, we get

\begin{equation}
A_n(k,\theta)=\underline{O} \left(\exp\left[\epsilon^{-d}
\|v_n\|_2^2\right]\right) \exp\left[ -\sum\limits_{j=0}^{n-1}
\kappa_j \right]\Bigl[A^0(\theta,k)+\nu_n\Bigr] \label{bound-1}
\end{equation}
with
\begin{equation}
|\nu_n|<\|v\|_2\exp(C\epsilon^{-d})\sum\limits_{j=1}^n
\exp(-\epsilon j+C\epsilon^{-d}j^{0.5} )<\|v\|_2
\exp(C\epsilon^{-d}) \label{bound0}
\end{equation}
and
\begin{equation}
A_n(k,\theta)= \exp\left[ -\sum\limits_{j=0}^{n-1} \kappa_j
\right]\underline{O} \left( \exp[C\epsilon^{-d}]\right)
\label{bounda}
\end{equation} which proves (\ref{asymp1}),
(\ref{asymp2}). Now, we need to show (\ref{asymp3}), otherwise that
would not be an asymptotical result. Indeed, fix $k$. Then, from
(\ref{bound-1}) and (\ref{bound0}),  we easily get (\ref{asymp3}) as
$\|v\|_2\to 0$.

Let us show that the a.c. spectrum of $H$ fills $\mathbb{R}^+$. We
will prove that the interval $I^2=\{k^2, k\in I\}$ supports the a.c.
component of the spectrum.   Following \cite{killip, denisov},
consider an isosceles triangle $T$ in $\Pi$ with the base equal to
$I$ and the adjacent angles both equal to $\pi/\gamma_1$,
$\gamma_1>d$ with $d$ from (\ref{bounda}). Then, for simplicity, fix
$f$-- nonzero $L^2$ spherically-symmetric function with compact
support within, say, a unit ball centered at origin. Clearly, we can
find some point $k_0$ within this triangle $T$ at which
$|A^0(f,k_0,\theta)|>C>0$ for $\theta\in \Sigma$. It follows just
from the analyticity of $A^0(f,k,\theta)$ in $k$ for fixed $\theta$
and spherical symmetry in $\theta$ for fixed $k$. Let us fix this
$k_0$. Then, consider a new potential $\hat{V}=\chi_{|x|>R}V(x)$
with $R$ large enough. By Rozenblum-Kato theorem \cite{rs3}, the
a.c. spectrum is not changed. Now, by (\ref{asymp3}), we can choose
$R$ large enough (that means $\|\hat{v}_n\|_2$ is small) to
guarantee that
\begin{equation}
|\tilde{A}_n(f,k_0,\theta)|>C>0 \label{est6}
\end{equation}
uniformly in $n$ and all $\theta\in \Sigma$. Then, we will prove
that $\sigma_f'(k^2)>0$ for a.e. $k\in I$, where $d\sigma_f(E)$ is
the spectral measure of $f$ with respect to operator
$\hat{H}=-\Delta+\hat{V}$. That will show that $I^2$ is in the
support of the a.c. spectrum of $H$. Since $I$ is arbitrary, that
will mean $\sigma_{ac}(H)=\mathbb{R}^+$.

To implement this strategy, we use the following factorization
identity (\cite{yafaev}, pages 40-42)

\begin{equation}
\sigma'_{n,f}(E)=k\pi^{-1}\|A_n(f,k,\theta)\|_{L^2(\Sigma)}^2, E=k^2
\label{factor}
\end{equation}
where $\sigma_{n,f}(E)$ is the spectral measure of $f$ with respect
to the operator with potential $\hat{V}_n=\chi_{|x|>R}V_n$ and $A_n$
is an amplitude with respect to the same potential. Let
$\omega(k_0,s), s\in \partial T$ denote the value at $k_0$ of the
Poisson kernel associated to $T$. One can easily show that

\begin{equation}
0\leq \omega(k_0,s)<C|s-s_{1(2)}|^{\gamma_1-1},\;  s\in
\partial T \label{bound}
\end{equation}
where $s_{1(2)}$ are endpoints of $I$. It is also nonnegative
function. Let us write the following inequalities
\[\int\limits_I \omega(k_0,s)
\ln\|A_n(f,s,\theta)\|^2_{L^2(\Sigma)}ds\]
\[
=\int\limits_I \omega(k_0,s) \ln \int_{\Sigma}
|WKB_n(s,\theta)\tilde{A}_n(f,s,\theta)|^2 d\theta ds>2(J_1+J_2)
\]
by Jensen's inequality, where
\[
J_1=\int\limits_I \omega(k_0,s)  \int_{\Sigma} \ln |WKB_n(s,\theta)|
d\theta ds
\]
\[
J_2=\int\limits_I \omega(k_0,s)  \int_{\Sigma} \ln
|\tilde{A}_n(f,s,\theta)| d\theta ds= \int_{\Sigma} \int\limits_I
\omega(k_0,s)\ln |\tilde{A}_n(f,s,\theta)|  dsd\theta
\]
We will estimate from below each of these terms. The function $\ln
|\tilde{A}_n (f,k,\theta)|$ is subharmonic in $k\in T$ for fixed
$\theta$, so we have a mean-value inequality
\[
\int\limits_I \omega(k_0,s) \ln|\tilde{A}_n(f,s,\theta)|ds\geq
\ln|\tilde{A}_n(f,k_0,\theta)|-\int\limits_{I_1\bigcup I_2}
\omega(k_0,s) \ln|\tilde{A}_n(f,s,\theta)|ds
\]
where $I_{1(2)}$ are the sides of triangle $T$. From (\ref{bounda}),
(\ref{est6}), and (\ref{bound}), we obtain the bound
\[
\int\limits_I \omega(k_0,s) \ln|\tilde{A}_n(f,s,\theta)|ds>C>-\infty
\]
uniformly in $n$ and $\theta\in \Sigma$. Here, the possible growth
of $\ln |\tilde A_n(f,s,\theta)|$ near the real line is compensated
by the zero of the kernel $\omega(k_0,s)$. Therefore,
$J_2>C>-\infty$ uniformly in $n$. Thus, we are left to show that the
same is true for $J_1$.
\[
J_1=-(4\pi)^{-1}\int\limits_I \omega(k_0,s) \int\limits_{\Sigma}
d\theta \int \frac{\cos(s(|u|-<\theta,u>))}{|u|}V_n(u)du
\]
\[
=-(4\pi)^{-1}\int\limits_I \omega(k_0,s) \int
\frac{\sin(2s|u|)}{s|u|^2} V_n(u)duds
\]
Since $\omega(k_0,s)$ is smooth on $I$ and equals to $0$ at the
endpoints, we have
\begin{equation}|J_1|<C\int |V_n(u)| (1+|u|^2)^{-1.5}du<C
\label{wkb4}
\end{equation} upon integration by parts in $s$.

Thus, from (\ref{factor}), we get
\[
\int\limits_I \ln \sigma'_{n,f}(k^2)dk>C
\]
uniformly in $n$. But then the standard argument on the
semicontinuity of the entropy (see \cite{kils}, Section 5) shows
that
\[
\int\limits_I \ln \sigma'_f(k^2) dk>-\infty
\]
It follows from the fact that $d\sigma_{n,f}(E)$ converges weakly to
$d\sigma_f(E)$ as $n\to\infty$. Let us collect the conditions on
sparseness that we used:
\[
2R_n<R_{n+1},\sigma_n=R_n^{3.5}e^{R_n}(R_{n+1}-R_n)^{-1}<e^{-n},
nR_n R_{n+1}^{-0.5}<e^{-2n}, n<R_{n+1}-R_n
\]
Obviously, the condition on $\sigma_n$ is the strongest one and we
can satisfy all of them by requiring
\[
R_{n+1}>e^{\alpha R_n}, \alpha>1
\]
and $R_0$ is big enough. In particular, $R_n=g^{(n)}(R_0)$ will work
for $g(x)=e^{2x}$.
\end{proof}
We believe that the restrictions on sparseness can be relaxed by
more detailed, rather straightforward analysis. It is also likely
that by controlling the oscillatory integrals for real $k$ one can
show that the spectrum is purely a.c. on $\mathbb{R}^+$. We do not
want to pursue that technically difficult problem in this paper. It
 also might be that formula $(5.4)$ from \cite{denisov} can be used
to obtain the asymptotics of Green's function in a simpler way.
Notice also that condition (\ref{barry}) is satisfied under the
assumptions of the theorem.

The $WKB$ asymptotics we proved in the theorem is quite new to the
best of our knowledge. It is different from the correction obtained
in \cite{perelman} and in earlier papers. Consider the randomized
model: e.g., $ v_n(x)=\sum_{k\in \Delta_n} \omega_k^n \nu_k^n(x)$
where $\{\omega_k^n\}$ are independent random variables with mean
zero and uniformly bounded dispersion, functions $\nu_k^n(x)$,
($k\in \Delta_n$-- set of indexes)-- bump functions living within
the small nonintersecting balls all lying inside the $n$--th
spherical layer, which also satisfy the bound $\max_{k\in \Delta_n}
\|\nu_k^n\|_\infty \in \ell^2(\mathbb{Z}^+)$ in $n$. Then,

\[
\mathbb{E}_\omega \left|\,\int\limits_{\mathbb{R}^3}
\frac{\exp(ik(|t|-<\theta, t>))}{|t|} V_n(t)dt\right|^2 <C
\]
uniformly in $n\in \mathbb{Z}^+$, $k\in \overline{\mathbb{C}^+}$,
and $\theta\in \Sigma$. So, one does not have any modification to
the asymptotics really. That is due to oscillations and was observed
before \cite{denisov}.

 Now, we want to discuss the following issue. In the last theorem,
 we  established the WKB asymptotics away from the real line. Recall that in the one-dimensional situation,
 the corresponding WKB correction was given by
 \[
 WKB(k)=\exp\left( -\frac{i}{2k} \int\limits_0^\infty V(r)dr
 \right)
\]
and its absolute value is equal to one for real $k$. Clearly, this
is not the case for the multidimensional WKB that we have got. We
had an estimate (\ref{wkb4}) that was sufficient to conclude the
presence of a.c. spectrum but rather than that this WKB can exhibit
quite a bad behavior. Apparently, the actual WKB asymptotics should
be understood differently. The level sets of the function
$|t|-<t,\theta>$ from the formula~(\ref{asymp2}) are paraboloids and
it suggests that some evolution equation of the heat-transfer type
might be involved. Consider operators $O_t$ given by the formula
(\ref{ooo}).

\begin{lemma}
The following parametrix representation is true for any $k\in
{\mathbb{C}^+}$
\begin{equation}
-2ik {O}_tf=\exp\left[-\frac{B}{2ikt}\right]f
+\underline{O}(t^{-1})\|f\|_2,\quad t>1 \label{parametrix}
\end{equation}
where $B$ is the Laplace-Beltrami operator on the unit sphere.
\end{lemma}
\begin{proof}
It is easy to show that $\displaystyle -2ik {O}_t\to I$ in the
strong sense as $t\to\infty$. Consider $\displaystyle
g_t=\exp\left[-\frac{B}{2ikt}\right]f$. It solves the following
problem
\[
g'=\frac{B}{2ikt^2}\,g,\, g(\infty)=f
\]
Take $\displaystyle \psi(t)=-2ik {O}_tf-g(t)$. Then,
\[
\psi'-\frac{B}{2ikt^2}\psi=C\left[ \, \int\limits_{\Sigma}
e^{ikt(1-<x,y>)}(1-<x,y>)f(y)dy+\right.
\]
\[
\left. +ikt\int\limits_{\Sigma} e^{ikt(1-<x,y>)}[1-<x,y>+\frac 12
(<x,y>^2-1)]f(y)dy\right]
\]
We estimate the integral operators in $L^{1,1}$, $L^{\infty,\infty}$
norms first and then interpolate by Riesz-Thorin theorem to get
\[
\psi'-\frac{B}{2ikt^2}\psi=\underline{O}(t^{-2})\|f\|_2,
\psi(\infty)=0
\]
Since $B$ is nonpositive, we get (\ref{parametrix}) by integration.
\end{proof}
Consider the following evolution equations
\begin{equation}
\frac{d}{dt}U_0(\tau,t,k)=-(2ik)^{-1}\frac{B}{t^2}U_0(\tau,t,k),
U_0(\tau,\tau,k)=I
\end{equation}
\begin{equation}
\frac{d}{dt}U(\tau,t,k)=-(2ik)^{-1}\left[\frac{B}{t^2}-V(t)\right]U(\tau,t,k),
U(\tau,\tau,k)=I
\end{equation}
Then,
\[
\exp\left[-\frac{B}{2ikt}\right]=U_0(t,\infty,k)
\]
 Since $A_{R_n}$ oscillates
relatively slow with the respect to the $(n+1)$-th layer, an
expression
\[
\int\limits_{R_{n+1}}^{R_{n+1}+1}O_{t}q_{t}dt
\]
from (\ref{iter}) basically coincides with the linear in $V$ term
for the Duhamel expansion of $U(R_{n+1}, R_{n+2},k)$.

An interesting open question is: what is the WKB correction for the
real $k$? The likely candidate might be a solution to the following
evolution problem
\begin{equation}
2iku_r=-\frac{B}{r^2}u(r,k)+V(r)u(r,k)
\end{equation}
 Unfortunately, we cannot control
the terms corresponding to the multiple collisions within the same
layer (e.g. higher order in $v_n$ terms) for the real $k$. But if
one considers only those that are linear in $v_n$, then the
conjecture seems to be reasonable. The same evolution equation can
be obtained via the formal asymptotical expansion for the 3-dim
Schr\"odinger operator written as the one-dimensional operator with
operator-valued potential. This new candidate for the correct WKB
and modification of wave operators preserves the $L^2(\Sigma)$ norm.
Notice that for $k$ real, we cannot reduce the asymptotics to the
scalar version because $v_{n+1}$ can be rough. So there is no any
contradiction really with what we proved in theorem \ref{theorem1}.

  Although we can prove asymptotics of Green's function for $k\in
\mathbb{C}^+$ only, some analysis is possible on the real line too.
The following simple result on the absence of embedded eigenvalues
holds
\begin{theorem}
If $R_n$ is sparse enough, then there are no positive eigenvalues
for any bounded $v_n$.
\end{theorem}
\begin{proof}
The idea is quite simple and was used before to treat the
one-dimensional problem: we show that the solutions can not decay
too fast between the layers where the potential is zero. Then, since
the solution and its gradient are square summable in $\mathbb{R}^3$,
we get the contradictions for $R_n$ large enough. To make the
argument work, we also need some apriori bounds from below that we
borrow from \cite{bk}.

 Assume that $\psi(x)$ is a real-valued eigenfunction corresponding
to an eigenvalue $E>0$. Since $\psi(x)\to 0$ as $|x|\to\infty$ (see
Chapter~2, \cite{cfks}), there is a point $x_0$ such that
$\psi(x_0)=\max_{x\in \mathbb{R}^3} |\psi(x)|=1$, this is our
normalization of $\psi$. We also know that $\|\psi\|_2$ is finite
but we have no control over this quantity.

Introduce the spherical change of variables and consider
$\omega(r)=r\psi(r\sigma)$, $\sigma\in \Sigma$, $x=r\sigma$. Now,
$\omega(r)\in L^2(\mathbb{R}^+,L^2(\Sigma))$. Expanding in the
spherical harmonics,
\[
\omega(r)=\sum\limits_{m=0}^\infty \sum\limits_{l_m=-m}^m
Y_{m,l_m}(\sigma) f_{m,l_m}(r)
\]
and we have for any $l_m$
\begin{equation}
-f''_{m,l_m}(r)-m(m+1)r^{-2}f_{m,l_m}(r)=Ef_{m,l_m}(r),\,
R_{n}+1<r<R_{n+1} \label{eq1}
\end{equation}
From  \cite{bk}, lemma 3.10, we infer the following bound
\begin{equation}
\int\limits_{R_n+1<|x|<R_{n}+2} \psi^2(x)dx>CR_n^2e^{-\gamma
R_n^{4/3}\ln R_n}, n>>1 \label{absi1}
\end{equation}
with $\gamma(E)>0$ which is an independent constant (at this point
we used the normalization of $\psi$ at $x_0$). Since the potential
$V$ is bounded, we have
\begin{equation} \|\Delta \psi\|_\infty<C, \|\nabla \psi\|_\infty <C \label{absi2}
\end{equation}
The bounds (\ref{absi1}), (\ref{absi2}) lead to the existence of
$r_n\in [R_n+1, R_n+2]$ such that
\begin{equation}
\sum\limits_{m=0}^\infty \sum\limits_{l_m=-m}^m
|f_{m,l_m}(r_n)|^2>CR_n^2 e^{-\gamma R_n^{4/3}\ln R_n}
\label{bound1}
\end{equation}
\begin{equation}
\sum\limits_{m=0}^\infty (m+1)^2 \sum\limits_{l_m=-m}^m
|f_{m,l_m}(r_n)|^2<CR_n^2 \label{bound2}
\end{equation}
Consider
\[
p_m=\sum\limits_{k=m}^\infty \sum\limits_{l_k=-k}^k
|f_{k,l_k}(r_n)|^2
\]
Then, (\ref{bound2}) leads to
\[
\sum\limits_{m=1}^{\infty} m p_m<CR_n^2
\]
and
\[
p_m<CR_n^2m^{-1}
\]
Take integer $k_n$ so large that $R_n^2k_n^{-1}<R_n\exp(-\gamma
R_n^{4/3}\ln R_n)$. From (\ref{bound1}), we have an estimate

\[
\sum\limits_{m=0}^{k_n}\sum\limits_{l_m=-m}^m
|f_{m,l_m}(r_n)|^2>(C_1R_n^2-C_2R_n)\exp(-\gamma R_n^{4/3}\ln R_n)
\]
with $C_1>0$.
 The following estimate easily follows from (\ref{eq1}) (e.g., by introducing the
Pr\"ufer transform \cite{kls}, formula $(2.4)$)
\[
|f_{m,l_m}(r)|^2+E^{-1}|f_{m,l_m}'(r)|^2
\]
\[
>\left[|f_{m,l_m}(r_n)|^2+E^{-1}|f_{m,l_m}'(r_n)|^2\right]\exp\left[
-(2\sqrt{E})^{-1}m(m+1)(r-r_n)r^{-1}r_n^{-1}\right]
\]
\[
>|f_{m,l_m}(r_n)|^2\exp\left[
-(2\sqrt{E})^{-1}(m+1)^2r_n^{-1}\right]
\]
Integrating the last estimate in $r$ and summing over the indices,
we have

\begin{equation}
C>\int\limits_{r_n<|x|<R_{n+1}}
[\psi^2(x)+E^{-1}(\psi_r(x)+r^{-1}\psi(x))^2] dx \label{contra2}
\end{equation}
\[
>(R_{n+1}-r_n) \sum\limits_{m=0}^\infty \sum\limits_{l_m=-m}^m
|f_{m,l_m}(r_n)|^2\exp\left[  -(2\sqrt{E})^{-1}(m+1)^2 r_n^{-1}
\right]
\]
\begin{equation}
>(R_{n+1}-R_n-2)(C_1R_{n}^2-C_2R_n)\exp(-\gamma R_n^{4/3}\ln
R_n)\exp\left[ -(2\sqrt{E})^{-1}(k_n+1)^2(R_n+2)^{-1}\right]
\label{contra}
\end{equation}
Now, choose $\{R_n\}$ so sparse that for any $\gamma>0$ and $E>0$
\[
(R_{n+1}-R_n-2)R_n^2 \exp\left[-\gamma R_n^{4/3}\ln
R_n-(2\sqrt{E})^{-1}(k_n+1)^2 (R_n+2)^{-1}\right]\to \infty, \,{\rm
as} \,n\to\infty
\]
Then, for $n$ large enough, we get the contradiction in
(\ref{contra}) because the constant in the left hand side of
(\ref{contra2}) is independent of $n$. Notice that the sparseness
conditions were chosen independent of the eigenfunction, eigenvalue
$E$, and even of $\|V\|_\infty$. It is satisfied, for instance, if
$R_{n+1}>\exp(\exp (R_n^\beta)), \beta>4/3$ and $R_0$ is large
enough.
\end{proof}
{\bf Remark.} Apparently, the estimates from \cite{bk} that we used
can be improved in our case. Thus, the sparseness conditions can
also be relaxed. \vspace{0.5cm}

{\bf Acknowledgements.} We are grateful to A. Kiselev, A. Seeger,
and R. Shterenberg for the stimulating discussions.


\begin{thebibliography}{99}

%\bibitem{sun} T. Aktosun, A factorization of the scattering matrix
%for the Sch\"{o}dinger equation and for the wave equation in one
%dimension, J. Math. Phys., Vol. 33, 1992, no. 11, 3865--3869.

\bibitem{bk} J. Bourgain, C. Kenig, On localization in the
continuous Anderson-Bernoulli model in the higher dimension,
Inventiones Mathematicae, Vol. 161, 2005, no. 2, 389--426.

\bibitem{cfks} H. Cycon, R. Froese, W. Kirsch, B. Simon, Schr\"odinger operators with application to quantum
mechanics and global geometry, Texts and Monographs in Physics,
Springer Study Edition, 1987.

\bibitem{denisov}
S.A. Denisov, Absolutely continuous spectrumm for multidimensional
Schr\"odinger operators, IMRN, 2004, no. 74, 3963--3982.

\bibitem{d1}
S.A. Denisov, On the preservation of absolutely continuous spectrum
for Schr\"odinger operators, to appear in J. Funct. Anal.


\bibitem{d2}
S.A. Denisov, A. Kiselev, Spectral properties of Schrodiinger
operators with decaying potentials, to appear in Festschrift for B.
Simon's 60-th birthday, Proceedings of Symposia in Pure Mathematics.

\bibitem{klein} F. Germinet, A. Klein, Operator kernel estimates for
functions of generalized Schr\"odinger operator, Proceedings of AMS,
Vol. 131, 2002, no.3, 911--920.




\bibitem{killip} R. Killip, Perturbations of one-dimensional Schr\"{o}dinger operators
preserving the absolutely continuous spectrum, IMRN, 2002, no. 38,
2029--2061.

\bibitem{kils}  R. Killip, B. Simon, Sum rules for Jacobi matrices and their
applications to spectral theory, Annals of Math., Vol. 158, 2003,
253--321.

\bibitem{kls} A. Kiselev, Y. Last, B. Simon, Modified Pr\"{u}fer
and EFGP transforms
  and the spectral analysis of one-dimensional Schr\"odinger
  operators,
  Comm. Math. Phys., Vol. 194,  1998,  no. 1, 1--45.

\bibitem{krishna} M. Krishna, Absolutely continuous spectrum for
sparse potentials, Proc. Indian Acad. Sci. Math. Sci., Vol. 103
1993, no. 3, 333--339.

\bibitem{lns} A. Laptev, S. Naboko, O. Safronov, A Szeg\"o condition for a
multidimensional Schr\"odinger operator, J. Funct. Anal., Vol. 219,
2005, no.2, 285-305.

\bibitem{lns1} A. Laptev, S. Naboko, O. Safronov, Absolutely continuous
spectrum of Schr\"odinger operators with slowly decaying and
oscillating potentials, Comm. Math. Phys., Vol. 253, 2005, no.3,
611-631.

\bibitem{molchanov} S. Molchanov, B. Vainberg, Spectrum of multidimensional
Schr\"odinger operators with sparse potentials, Analytical and
computational methods in scattering and applied mathematics (Newark,
DE, 1998), 231--254, Chapman and Hall/CRC Res. Notes Math., 417,
2000.

\bibitem{mv} S. Molchanov, B. Vainberg, Scattering on the system of sparse bumps: multidimensional case,
Applicable Analysis, Vol. 71, 1999, 167--185.

\bibitem{pearson} D. B. Pearson, Singular continuous measures
in scattering theory, Comm. Math. Phys., 60, 1978, no. 1, 13--36.

\bibitem{perelman}
G. Perelman, Stability of the absolutely continuous spectrum for
multidimensional Schr\"odinger operators, IMRN,  2005, no. 37,
2289--2313.


\bibitem{rs3} M. Reed, B. Simon, Methods of modern
mathematical physics. III. Scattering theory. Academic Press, 1979.

\bibitem{oleg} O. Safronov, On the a.c. spectrum of
multi-dimensional Schr\"odinger operators with slowly decaying
potentials, Comm. Math. Phys., Vol. 254, 2005, no. 2, 361--366.



\bibitem{simon} B. Simon, Schr\"odinger operator in the 21-st century, Imp.
Coll. Press, London, 2000, 283--288.

%\bibitem{stein} E. Stein, Harmonic analysis: real varaible methods,
%orthogonality, and oscillatory integrals, Princeton University
%Press, 1993.

\bibitem{yafaev} D. Yafaev,
``Scattering theory: some old and new problems",
 Lecture Notes in Mathematics, 1735. Springer-Verlag, Berlin,
 2000.


\end{thebibliography}
\end{document}